\documentclass[%
superscriptaddress,
preprint,
 amsmath,amssymb,
 aps,
]{revtex4-1} 

\usepackage{graphicx}
\usepackage{dcolumn}
\usepackage{bm}
\usepackage{xcolor}

\begin{document}


\title{Superluminal matter waves}

\author{J.P. Palastro}
\email{jpal@lle.rochester.edu}
\affiliation{
University of Rochester, Laboratory for Laser Energetics, Rochester, New York 14623-1299 USA}

\author{D. Ramsey}
\affiliation{
University of Rochester, Laboratory for Laser Energetics, Rochester, New York 14623-1299 USA}

\author{M. Formanek}
\affiliation{ELI Beamlines Facility, The Extreme Light Infrastructure ERIC, 252 41 Doln\'{i} B\v{r}e\v{z}any, Czech Republic}

\author{J. Vieira}
\affiliation{GoLP/Instituto de Plasmas e Fusão Nuclear, Instituto Superior Técnico, Universidade de Lisboa, Lisbon 1049-001, Portugal}

\author{A. Di Piazza}
\affiliation{
University of Rochester, Laboratory for Laser Energetics, Rochester, New York 14623-1299 USA}
\affiliation{Department of Physics and Astronomy, University of Rochester, Rochester, New York 14627, USA}

\date{\today}

\begin{abstract}
The Dirac equation has resided among the greatest successes of modern physics since its emergence as the first quantum mechanical theory fully compatible with special relativity. This compatibility ensures that the expectation value of the velocity is less than the vacuum speed of light. Here, we show that the Dirac equation admits free-particle solutions where the peak amplitude of the wavefunction can travel at any velocity, including those exceeding the vacuum speed of light, despite having a subluminal velocity expectation value. The solutions are constructed by superposing basis functions with correlations in momentum space. These arbitrary velocity wavefunctions feature a near-constant profile and may impact quantum mechanical processes that are sensitive to the local value of the probability density as opposed to expectation values.
\end{abstract}
                     
\maketitle
The discovery of the Dirac equation stands as a foundational achievement of modern theoretical physics \cite{Dirac1928}. As the first quantum mechanical theory fully compatible with special relativity, the Dirac equation resolved several inconsistencies that beset the, otherwise widely successful, Schr\"odinger equation \cite{Schrodinger1926}. The equation and its solutions preserve many of the quantum mechanical concepts developed within the context of the Schr\"odinger equation, such as probability currents, expectation values, and operators. Unlike the Schr\"odinger equation, however, the Dirac equation was formulated from the outset to exhibit a Lorentz-invariant and Hamiltonian structure consistent with special relativity. In doing so, the Dirac equation precluded unphysical phenomena fully allowed by the Schr\"odinger equation: namely, expectation values for velocity that exceed the vacuum speed of light.

At a fundamental level, the Dirac equation is a wave equation. Within first quantization, the solutions, or wavefunctions $\psi$, describe the quantum mechanical state of a charged lepton, while their adjoint product $\psi^{\dagger}\psi$ provides the probability density of finding the lepton within a particular region of space--time or momentum--energy. In the absence of fields (and vacuum nonlinearities), the simplest solution is a plane wave modulating a constant spinor \cite{peskin1995}. Physically occurring, localized wavefunctions are typically formed by superposing these solutions with amplitudes and phases that are uncorrelated in space--time or momentum--energy. Superpositions featuring correlations in space--time or momentum--energy allow for wavefunctions with more-complex and potentially advantageous structures.

Insight into the fundamental properties of matter and light and the potential for applications has driven a growing interest in structured solutions for quantum mechanical and electromagnetic waves, with ideas from one often being adapted to the other \cite{Berry1979,Allen1992,Berry1998,Bliokh2007,Verbeeck2010, Kaminar2012,Voloch-Bloch2013,Grillo2014, Kaminer2015,Loyd2017,Hall2023, Campos2024,Mihaila2022,Wong2024}. One of the most-iconic examples is the concept of orbital angular momentum from quantum mechanics being adapted to electromagnetism \cite{Allen1992,Berry1998}. Within quantum mechanics, this ``wavefunction engineering'' has been applied to: the formation of self-accelerating solutions, which appear to move under the influence of a potential in the absence of any potential \cite{Kaminer2015,Campos2024}; transverse shaping of electron wavefunctions for added flexibility in scanning electron microscopes \cite{Mihaila2022}; and the formation of free-electron crystals for enhancing x-ray radiation \cite{Wong2024}. In each of these, the physical processes of interest are sensitive to the local properties of the wavefunction as opposed to expectation values.

Here, we introduce a structured free-particle solution to the Dirac equation where the peak amplitude of the wavefunction can travel at any velocity, including those exceeding the vacuum speed of light, despite having a subluminal velocity expectation value (Fig. 1). Motivated by similar solutions found in electromagnetism (e.g., space--time wavepackets \cite{Longhi03,Kondakci2017,Yessenov2022} and flying focus pulses \cite{Sainte-Marie2017,Froula2018,Palastro2018,DiPiazza2021, Formanek2023,Ramsey2023,Palastro2024}), the solutions feature a near-constant profile and are constructed by superposing basis functions with correlations in momentum--energy.  We expect that the solutions, or approximations thereof, could be produced by light-matter interactions, such as the Kapitza--Dirac effect \cite{Kapitza1933,Freimund2002, Shiloh2014,Grillo2014,Vanacore2018,Feist2020,Reinhardt2020,Abajo2021, Lin2024}, and generalized to exhibit more exotic properties, such as a modified effective mass in the absence of fields. These arbitrary velocity wavefunctions may impact quantum mechanical phenomena that are sensitive to the local value and velocity of the probability density, like Smith--Purcell or Cherenkov radiation \cite{Pupasov2021}, as opposed to expectation values.

The wavefunction of a spin one-half charged particle evolves according to the Dirac equation. In natural units ($\hbar = c = 1$), the Schr\"odinger form of the Dirac equation is given by
\begin{equation}\label{eq:SchDirac}
i \partial_t \psi = H\psi,
\end{equation}
where $H = -i\bm{\alpha} \cdot \bm{\nabla} + \beta m$ is the Hamiltonian, 
\begin{align}
\alpha^{j} = \begin{pmatrix}   
          0 &  \sigma^j \\
          \sigma^j & 0
          \end{pmatrix}, && \hspace{-40pt} 
\beta = \begin{pmatrix}   
          I &  0 \\
          0 & -I
          \end{pmatrix},  \notag
\end{align}
the $\sigma^j$ with $j = 1,2,3$ are the Pauli matrices, $I$ is the identity matrix, and $m$ is the mass of the particle. Here and throughout, bold font denotes three-vectors, $\bm{a}^2 \equiv \bm{a} \cdot \bm{a}$, the shorthand notations $p \equiv p^{\mu}$ and $x \equiv x^{\mu}$ are used for the momentum and position four-vectors, $a\cdot b \equiv a^{\mu}b_{\mu} = a_0b_0 - \bm{a}\cdot \bm{b}$, and relativistic notation for sub and superscripts is not used.

The Dirac equation admits four plane wave solutions, corresponding to positive and negative energy and spin states. A general wavefunction can be expressed as a superposition of these solutions. For simplicity and definiteness, a superposition of solutions with positive energy and spin in the rest frame will be considered, such that 
\begin{equation}\label{eq:psi}
\psi(x) = \frac{1}{(2\pi)^{3/2}}\int \frac{1}{\sqrt{2p_0}} u(p) f(p) e^{-ip\cdot x} \delta(p_0 - E_{\bm{p}})d^4p,
\end{equation}
where 
\begin{equation}\label{eq:f}
u(p) = \sqrt{\scriptstyle E_{\bm{p}} + m} 
\begin{pmatrix}  
      \chi_+  \\ 
          \frac{\bm{\sigma} \cdot \bm{p}}{E_{\bm{p}} + m} \chi_+ \\
          \end{pmatrix},
\end{equation}
is the bispinor normalized so that $u^{\dagger}\beta u = 2m$, $\dagger$ indicates a Hermitian conjugate, $\chi_+ = (\begin{smallmatrix}1 \\ 0 \end{smallmatrix})$, $E_{\bm{p}}=(m^2 + \bm{p}^2 )^{1/2} > 0$, and the Dirac delta function $\delta$ ensures the ``on-shell'' condition. The complex scalar function $f(p)$ determines the relative phase and amplitude of each plane wave that composes the wavefunction. Here, $f(p)$ is expressed as an integral over an auxiliary parameter $q$:
\begin{equation}\label{eq:f}
f(p) =  \int N(p,q) dq.
\end{equation}
The utility of this auxiliary parameter will become clear below. The functions $f(p)$ and $N(p,q)$ are constrained by the normalization condition
\begin{equation}
\int \psi^{\dagger}\psi d^3x = \int |f(p)|^2 d^3p = 1,
\end{equation}
where $\psi^{\dagger}\psi$ is the probability density and, for the remainder, $p_0 = E_{\bm{p}}$ is implied.

The evolution of the wavefunction can be characterized by three types of velocities: the phase velocities of the plane wave solutions, the eigenvalues (or expectation values) of the velocity operator, and the group velocity. The phase velocities, $\bm{v}_p = E_{\bm{p}}\bm{p}/\bm{p}^2$, are always superluminal ($|\bm{v}_p| >1$). The velocity operator $\bm{v}$ is found by evaluating the commutator of the Hamiltonian and position vector:  
$\bm{v} = i[H,\bm{x}] = \bm{\alpha}$. Using Eq. \eqref{eq:psi}, one finds $\langle\bm{v}\rangle = \int \psi^{\dagger} \bm{v} \psi d^3x =  \int (\bm{p}/E_{\bm{p}})|f(p)|^2 d^3p = \langle\bm{p}/E_{\bm{p}}\rangle$, which, as expected, is always subluminal ($ |\bm{p}/E_{\bm{p}}| < 1$).

The group velocity $\bm{v}_g$ can take any value in any direction. For a typical wavefunction, the components of the four momenta must satisfy the on-shell condition but are otherwise independent. As a result, $\partial p_j/\partial p_k = \delta_{jk}$ and
\begin{equation}
v_{g,j} = \frac{ \partial E_{\bm{p}}}{\partial p_j} = \frac{p_j}{E_{\bm{p}}}, 
\end{equation}
which is consistent with the eigenvalues of the velocity operator $\langle\bm{v}\rangle = \langle\bm{p}/E_{\bm{p}}\rangle$. However, if the momenta have some correlation, i.e., they are interdependent, then $\partial p_j/\partial p_k \neq \delta_{jk}$ and
\begin{equation}\label{eq:newvg}
v_{g,j} = \frac{ \partial E_{\bm{p}}}{\partial p_j} = \frac{p_j}{E_{\bm{p}}} + \frac{1}{2E_{\bm{p}}}\frac{ \partial }{\partial p_j}(\bm{p}^2 - p_j^2). 
\end{equation}
In this case, the correlated momenta introduce an additional contribution (the second term on the right-hand side) that decouples the group velocity from $\bm{p}/E_{\bm{p}}$. 

\begin{figure}
\includegraphics[width=1\linewidth]{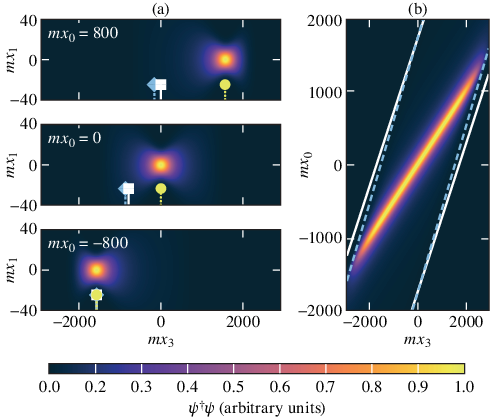}
\caption{Evolution of the probability density $\psi^\dagger \psi$ for a lepton with a velocity expectation value $\langle v_{3} \rangle = 0.9$ and a group velocity $v_a = 2$. The peak of the probability density travels superluminally while maintaining a near-constant profile. The yellow (circle, dotted), white (square, solid), and blue (diamond, dashed) lines indicate signals travelling at the velocities 2, 1, and 0.9, respectively. In this example, $\bar{\mathcal{P}} = 2m$, $w = 0.1m$, $\xi_0 = 250 m^{-1}$, and $\Delta \zeta = 1400 m^{-1}$ [see Eqs. \eqref{eq:parsol} -- \eqref{eq:psiapp}]. }
\label{fig:f1}
\end{figure}

\begin{figure}
\includegraphics[width=1\linewidth]{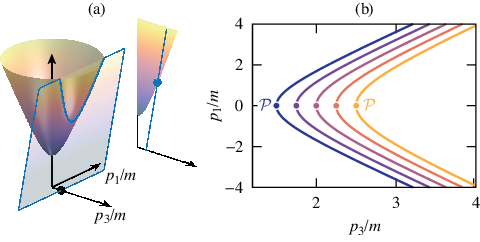}
\caption{(a) Satisfying the on-shell condition for an arbitrary group velocity $v_a$ is equivalent to finding the intersection between a four-dimensional hyperboloid, $m^2 = E_p^2 - \bm{p}^2$, and the hyperplane 
$E_{\bm{p}} = \bm{v}_a \cdot \bm{p} + \kappa.$ Here, the dimensionality has been reduced for visualization purposes by setting $p_2 = 0$. (b) Projections of the intersection in the $p_3$--$p_1$ plane for different values of $\mathcal{P}$. The values of $\mathcal{P}$ define the vertices of the projection, i.e., the values of $p_3$ where $p_1 = p_2 = 0$ [Eq. \eqref{eq:ppar2}]. The parameters are the same as in Fig. 1.}
\label{fig:f2}
\end{figure}

As an example, consider the case of an arbitrary, constant group velocity $\bm{v}_a$. Direct integration of $\partial E_{\bm{p}} / \partial p_j$ yields
\begin{equation}\label{eq:Epva}
E_{\bm{p}} = \bm{v}_a \cdot \bm{p} + \kappa,
\end{equation}
where $\kappa$ is a constant. Setting $ (m^2 + \bm{p}^2 )^{1/2} = \bm{v}_a \cdot \bm{p} + \kappa$ ensures that the on-shell condition is satisfied and determines the interdependence of the momenta (Fig. \ref{fig:f2}). Specifically, 
\begin{equation}\label{eq:pcond}
\bm{p}^2 - (\bm{v}_a \cdot \bm{p})^2 - 2\kappa(\bm{v}_a \cdot \bm{p}) + m^2 - \kappa^2 = 0.
\end{equation}
Without loss of generality, the specified group velocity $\bm{v}_a$ will be aligned along the $j=3$ axis, i.e., $\bm{v}_a = v_a \bm{e}_3$, where $\bm{e}_j$ denotes a unit vector. Substitution of Eqs. \eqref{eq:Epva} and \eqref{eq:pcond} into the right-hand-side of Eq. \eqref{eq:newvg} verifies that indeed $v_{g,3}  = v_a$. The condition on the momenta [Eq. \eqref{eq:pcond}] is a quadratic equation whose solution provides $p_3$ in terms of $\bm{p}_{\perp} \equiv p_1 \bm{e}_1 + p_2 \bm{e}_2$. Upon solving this equation, one finds
\begin{equation}\label{eq:ppar}
p_3 = p_c \equiv \kappa \gamma_a^2 v_a \pm \left[(\kappa  \gamma_a^2)^2 - \gamma_a^2(m^2 + \bm{p}_{\perp}^2) \right]^{1/2},
\end{equation}
where $\gamma_a \equiv 1/\sqrt{1-v_a^2}$. Note that $\gamma_a$ only appears in even powers. Thus, a superluminal group velocity ($|v_a| > 1$) does not produce a complex valued $p_c$. 

The interdependence of the momenta, as described by Eq. \eqref{eq:ppar}, can be built into the wavefunction by making use of the auxiliary parameter $q$. Expressing $q = q(\kappa)$ as a function of $\kappa$ (to be determined below) and setting
\begin{equation}\label{eq:N}
N(p,q) = \mathcal{N}(p,q)\delta(p_3 - p_c)
\end{equation}
yields
\begin{equation}\label{eq:unew}
f(p) = \int \mathcal{N}(p,q) \delta(p_3 - p_c)dq ,
\end{equation}
where the dependence of $p_c$ and $q$ on $\kappa$ is understood. Together, Eqs. \eqref{eq:psi} and \eqref{eq:unew} describe a superposition of solutions with correlated momenta parameterized by $\kappa$. Each solution has the same arbitrary group velocity $v_{g,3} = v_a$. 

To elucidate the physical meaning of the parameter $\kappa$ and facilitate further calculation, it is helpful to substitute Eq. \eqref{eq:unew} into Eq. \eqref{eq:psi} and define the function
\begin{equation}\label{eq:unew2}
\begin{aligned}
\Phi(x,q) = \int & \frac{1}{\sqrt{2E_{\bm{p}}}} u(p) \mathcal{N}(p,q) e^{-ip\cdot x}  \delta(p_3 - p_c) d^3p,
\end{aligned}
\end{equation}
such that $\psi(x) = (2\pi)^{-3/2}\int \Phi(x,q) dq$. Several properties of the arbitrary group velocity solutions can be analyzed by considering the phase in Eq. \eqref{eq:unew2}: $p \cdot x = p_0x_0 - p_3x_3 - \bm{p}_{\perp}\cdot \bm{x}_{\perp}$. Making use of the delta function, the phase contribution $\phi \equiv p_0x_0 - p_3x_3 $ becomes
\begin{align}\label{eq:phi}
\phi &= (p_3v_a + \kappa )x_0 - p_3x_3  
= \kappa  x_0 + p_3(v_a x_0 - x_3).
\end{align}
The first term is proportional to $\kappa$ but does not depend on $p$. As a result, the factor $e^{-i\kappa  x_0}$ can be extracted from the integrand. This factor shows that $\kappa$ quantifies the temporal phase advance of the wavefunction. The second term in Eq. \eqref{eq:phi} reveals that the integrand depends on $x_0$ and $x_3$ only in the combination $v_ax_0 - x_3$. Thus, the integrand advects at the group velocity $v_a$.

An intuitive, and often helpful, picture of a wavefunction is that of an ``envelope'' modulated by a ``carrier'' wave. The envelope describes the bulk motion of the probability density and propagates at the group velocity, while the carrier describes the motion of the phase fronts, which propagate at the phase velocity. The function $\Phi(x,q)$ can be expressed in this same framework. Adding and subtracting a longitudinal (along $x_3$) momentum offset $\mathcal{P}$ to Eq. \eqref{eq:phi} provides
\begin{equation}\label{eq:phi2}
\phi = \mathcal{P}\left[(v_a + \tfrac{\kappa }{\mathcal{P}})x_0-x_3 \right] + (p_3-\mathcal{P})(v_ax_0  - x_3). 
\end{equation}
With this offset, the first and second terms in $\phi$ describe the phase of the carrier wave, which can be extracted from the integrand as before, and the advection of the integrand (i.e., the envelope) at the group velocity. Choosing
\begin{equation}\label{eq:kappa}
\kappa  = \mathcal{E}-v_a\mathcal{P},
\end{equation}
where $\mathcal{E} = (m^2 + \mathcal{P}^2)^{1/2}$, yields
\begin{equation}\label{eq:phi3}
\phi = \mathcal{P}(\beta_px_0-x_3) + (p_3-\mathcal{P})(v_ax_0  - x_3) 
\end{equation}
and ensures that the phase fronts of the carrier wave travel at a phase velocity $\beta_p \equiv \mathcal{E/P}$. Note that this is equivalent to the procedure of enveloping the wavefunction about the longitudinal momentum $\mathcal{P}$ and energy $\mathcal{E}$.

With the auxiliary parameter $\kappa  = \mathcal{E}-v_a\mathcal{P}$, the condition on the interdependence of the momenta [Eq. \eqref{eq:ppar}] simplifies to 
\begin{equation}\label{eq:ppar2}
p_c - \mathcal{P} = \gamma_a^2(v_a\mathcal{E} - \mathcal{P}) - \varpi\left[ \gamma_a^4(v_a\mathcal{E} - \mathcal{P})^2 - \gamma_a^2 \bm{p}_{\perp}^2 \right]^{1/2},
\end{equation}
where the choice of sign $\varpi \equiv \mathrm{sign}[\gamma_a^2(v_a\mathcal{E} - \mathcal{P})]$ ensures that the envelope varies slowly compared to the phase fronts. Equation \eqref{eq:ppar2} demonstrates two important points: First and foremost that $\mathcal{P}$ corresponds to the value of $p_3$ when $ \bm{p}_{\perp} =0$. And second that arbitrary group velocity solutions are precluded in one dimension. In one dimension, the wavefunction is composed solely of plane waves with $\bm{p}_{\perp} = 0$, such that $p_c$ is always equal to $\mathcal{P}$. This eliminates the second term in Eq. \eqref{eq:phi3}, which is responsible for the movement of the envelope at $v_a$. Figure \ref{fig:f2}(b) displays a family of longitudinal momenta $p_3$ [Eq. \eqref{eq:ppar2}] parameterized by $\mathcal{P}$ as a function of perpendicular momentum $|\bm{p}_{\perp}|$ for a superluminal group velocity $v_a = 2 $. 

A distinguishing property of the Dirac equation is its Lorentz covariance. Lorentz transformations of the arbitrary group velocity solutions [Eqs. \eqref{eq:unew2} and \eqref{eq:ppar2}] provides additional insight into their interpretation. Upon performing a longitudinal Lorentz transformation to a frame moving at a velocity $v_L$ with respect to the nominal frame, Eq. \eqref{eq:phi3} becomes
\begin{align}\label{eq:phiT}
\phi &= \frac{\mathcal{P}}{\gamma_L(1+\beta_p'v_L)} (\beta_p'x'_0-x'_3) \\ &\hspace{16pt}
 + \gamma_L(p_3-\mathcal{P})[(v_a-v_L)x'_0  - (1-v_av_L)x'_3], \notag
\end{align}
where $'$ denotes coordinates in the moving frame and $\beta'_p = (\beta_p - v_L)/(1-\beta_pv_L)$ is the nominal phase velocity in the moving frame. One may expect that $v_L = v_a$ would be a natural choice for the frame velocity. However, special care must be taken to ensure that the nominal phase velocity is always superluminal (i.e., the on-shell condition is satisfied). For $|v_a| < 1$, one can indeed set $v_L = v_a$ to find
\begin{align}\label{eq:phi<}
\phi_< = \frac{\mathcal{P}}{\gamma_a(1+\beta_p'v_a)} (\beta_p'x'_0-x'_3) - \frac{(p_3-\mathcal{P})}{\gamma_a}x'_3,  
\end{align}
but for $|v_a| > 1$ the on-shell condition requires $v_L = v^{-1}_a$, which yields
\begin{align}\label{eq:phi>}
\phi_> = \frac{v_a\mathcal{P}}{\gamma_a(v_a+\beta_p')} (\beta_p'x'_0-x'_3) - \frac{(p_3-\mathcal{P})}{v_a\gamma_a}x'_0.
\end{align}
Recall that the second terms in Eqs. \eqref{eq:phi<} and \eqref{eq:phi>} determine the space--time dependence of the envelope. Thus, the subluminal and superluminal solutions represent Lorentz transformations of solutions with time-independent and longitudinal-coordinate-independent envelopes, respectively.

The phase [Eq. \eqref{eq:phi3}] and momentum condition [Eq. \eqref{eq:ppar2}] appearing in Eq. \eqref{eq:unew2} depend directly on $\mathcal{P}$. Thus, a convenient choice for the auxiliary parameter is $q(\kappa) = \mathcal{P}(\kappa)$, or more explicitly $\mathcal{P}(\kappa) = \gamma_a^2v_a\kappa \pm (\gamma_a^4 \kappa^2 - \gamma_a^2m^2)^{1/2}$ [see Eq. \eqref{eq:kappa}]. Using this choice and applying the delta functions in Eq. \eqref{eq:unew2} provides
\begin{equation}\label{eq:unew3}
\begin{aligned}
\Phi(x,\mathcal{P}) &= e^{-i\mathcal{P}\eta} \int d\bm{p}_{\perp}  \tilde{\mathcal{N}}(\bm{p}_{\perp},\mathcal{P}) \\ &\begin{pmatrix} 
          \chi_+  \\ 
          \dfrac{\sigma^3 p_c + \bm{\sigma}_{\perp} \cdot \bm{p}_{\perp} }{v_a(p_c-\mathcal{P}) +\mathcal{E} + m} \chi_+ \\
          \end{pmatrix} 
          e^{-i(p_c-\mathcal{P})\xi + i\bm{p}_{\perp}\cdot \bm{x}_{\perp} }  ,
\end{aligned}
\end{equation}
where $\eta = \beta_px_0 - x_3$, $\xi = v_ax_0 - x_3$, 
\begin{equation}\notag
\tilde{\mathcal{N}}(\bm{p}_{\perp},\mathcal{P}) = \sqrt{\tfrac{\scriptstyle v_a(p_c-\mathcal{P}) +\mathcal{E}  + m}{2[\scriptstyle v_a(p_c-\mathcal{P}) +\mathcal{E}]}} \int \mathcal{N}(p,\mathcal{P})\delta(p_3 - p_c) dp_3,
\end{equation}
and the $\mathcal{P}$ dependence of $\beta_p$, $\mathcal{E}$, and $p_c$ are understood. As before, the integral (envelope) depends solely on the space--time coordinates in the combination $\xi = v_ax_0 - x_3$ and advects at $v_a$.

Approximate analytical solutions for $\Phi(x,\mathcal{P})$, and ultimately $\psi(x)$, can be found for a particle moving predominantly in the $\bm{e}_3$ direction with a velocity $\langle v_3 \rangle$ that is sufficiently distinct from the group velocity $v_a$. More specifically, one assumes $|v_a\beta_p - 1|\mathcal{P}^2 \gg w^2$, where $w$ is the transverse momentum spread. This condition describes a ``paraxial'' limit in which Eq. \eqref{eq:ppar2} simplifies to
\begin{equation}\label{eq:ppar3}
p_c  \approx  \mathcal{P} +  \frac{\bm{p}_{\perp}^2}{2\mathcal{P}(v_a\beta_p - 1)}.
\end{equation}
With the quadratic dependence of $p_c$ on $\bm{p}_{\perp}$, a natural choice of basis functions for $\tilde{\mathcal{N}}(\bm{p}_{\perp},\mathcal{P})$ are the Laguerre--Gaussian modes:
\begin{equation}\label{eq:Adecom}
\tilde{\mathcal{N}}(\bm{p}_{\perp},\mathcal{P}) = \sum_{n,\ell} \mathcal{A}_{n\ell}(\mathcal{P})\rho^{|\ell|}L_n^{|\ell|}(\rho^2)\mathrm{exp}(-\tfrac{1}{2}\rho^2)e^{i\ell\theta_p},
\end{equation}
where $\rho = w^{-1}(\bm{p}_{\perp}^2)^{1/2}$, $L_n^{|\ell|}$ is a generalized Laguerre polynomial, and $\theta_p = \mathrm{atan}(p_2/p_1)$ is the azimuth in momentum space. Upon substituting Eqs. \eqref{eq:ppar3} and \eqref{eq:Adecom} into Eq. \eqref{eq:unew3}, dropping terms higher order than $\bm{p}_{\perp}^2$ in the spinor, and integrating, one finds
\begin{equation}\label{eq:parsol}
\begin{aligned}
\Phi(x,\mathcal{P}) &\approx e^{-i\mathcal{P}\eta}  \sum_{n,\ell}   \tilde{\mathcal{A}}_{n\ell}(\mathcal{P})  
\left(\begin{smallmatrix} 
          U_1(x)  \\ 
          U_2(x) \\
          U_3(x) \\
          U_4(x) 
\end{smallmatrix}\right), 
\end{aligned}
\end{equation}
where $\tilde{\mathcal{A}}_{n\ell}(\mathcal{P}) =  2\pi i^{\ell} (-1)^n \mathcal{A}_{n\ell}(\mathcal{P})$. Each component of the spinor can be expressed in terms of $U_1$ as follows:
\begin{equation}\label{eq:bigU}
\begin{aligned}
U_1(x) &= \tfrac{w}{R}\left(\tfrac{r}{R}\right)^{|\ell|}L_n^{|\ell|}(\tfrac{r^2}{R^2})\mathrm{exp}\big[-(1-i\tfrac{\xi}{\xi_0})\tfrac{r^2}{2R^2} + i\Lambda(x) \big] \\ 
 U_2(x) &= 0 \\
 U_3(x) &= \tfrac{\mathcal{P}}{\mathcal{E} +m}\left[1-\tfrac{(\mathcal{E}-v_a\mathcal{P} + m)}{2w^2\xi_0(\mathcal{E}+m)\mathcal{P}}\nabla_{\perp}^2 \right] U_1(x) \\ 
 U_4(x) &= -\tfrac{i}{(\mathcal{E} +m)}r^{|\ell|}e^{i\theta}\partial_r r^{-|\ell|} U_1(x),
\end{aligned}
\end{equation}
where $r = (\bm{x}_{\perp}^2)^{1/2}$, $\theta = \mathrm{atan}(x_2/x_1)$ is the azimuth in configuration space, $R = w^{-1}[1+(\xi/\xi_0)^2]^{1/2}$, ${\Lambda(x) = \ell\theta -(2n+|\ell| +1)\mathrm{atan}(\tfrac{\xi}{\xi_0})}$, and
\begin{equation}\label{eq:xi0}
 \xi_0 = \frac{\mathcal{P}(v_a\beta_p - 1)}{w^2}.
\end{equation}
Equations \eqref{eq:parsol} -- \eqref{eq:xi0} show that the envelope advects at the group velocity $v_a$ while maintaining a constant profile characterized by the duration $\xi_0/v_a$.

The $\mathcal{A}_{n\ell}(\mathcal{P})$ quantify the distribution of $\mathcal{P}$ values for every $n$ and $\ell$ mode that compose the wavefunction. Said differently, the wavefunction is comprised of solutions that fall along a curve defined by Eq. \eqref{eq:ppar3} and parameterized by $\mathcal{P}$ [Fig. \ref{fig:f2}(b)]: $\psi(x) = (2\pi)^{-3/2}\int \Phi(x,\mathcal{P}) d\mathcal{P}$. If the $\mathcal{A}_{n\ell}(\mathcal{P})$ are narrowly peaked about some $\mathcal{P} = \bar{\mathcal{P}}$, an approximate expression for this integral and the wavefunction can be obtained. Expressing $\Phi(x,\mathcal{P}) = \sum_{n,\ell} \Phi_{n\ell}(x,\mathcal{P})$ yields 
\begin{equation}\label{eq:psiapp}
\begin{aligned}
\psi(x) \approx  \sum_{n,\ell} \frac{ \Phi_{n\ell}(x,\bar{\mathcal{P}}) }{ \tilde{\mathcal{A}}_{n\ell}(\bar{\mathcal{P}}) } \int \tilde{\mathcal{A}}_{n\ell}(\bar{\mathcal{P}} + \Delta \mathcal{P})e^{-i\zeta \Delta \mathcal{P}} d \Delta \mathcal{P},
\end{aligned}
\end{equation}
where $\Delta \mathcal{P} = \mathcal{P} - \bar{\mathcal{P}} $, $\zeta = v_n x_0 - x_3$, and $v_n = \bar{\mathcal{P}}/(m^2 + \bar{\mathcal{P}}^2)^{1/2}$. Figure \ref{fig:f1} displays the components of a  wavefunction with a superluminal velocity $v_a = 2$ for $n = \ell = 0$ and $\int \tilde{\mathcal{A}}_{00}(\bar{\mathcal{P}} + \Delta \mathcal{P})e^{-i\zeta\Delta \mathcal{P} } d \Delta \mathcal{P} \propto \mathrm{exp}[-(\zeta/\Delta \zeta)^8]$.

Solutions to the Dirac equation can feature a peak probability density that  moves at any velocity, including those exceeding the speed of light, while maintaining a near-constant profile. This motion is independent of the velocity expectation value. The solutions are on-shell and constructed by superposing basis functions with correlations in the longitudinal and transverse momenta. Future work will investigate electromagnetic structures that can produce these wavefunctions through the Kapitza--Dirac effect; explore the benefit of these wavefunctions in phenomena, such as Compton scattering and Smith--Purcell and Cherenkov radiation; and pursue additional structures by replacing Eq. \eqref{eq:Epva} with other dependencies, like $E_{\bm{p}} = p_1^2/\mathcal{M} + \kappa$, which could exhibit classical dispersion with a modified effective mass $\mathcal{M}$ in vacuum. 

\begin{acknowledgments}
The authors would like to thank R. R. Almeida and D. H. Froula for insightful discussions. The work of M.F. is supported by the European Union’s Horizon Europe research and innovation program under the Marie Skłodowska-Curie Grant Agreement No. 101105246-STEFF. The work of J.P.P., D.R., and A.D. is supported by the Office of Fusion Energy Sciences under Award Numbers DE-SC0021057, the Department of Energy National Nuclear Security Administration under Award Number DE-NA0004144, the University of Rochester, and the New York State Energy Research and Development Authority. This report was prepared as an account of work sponsored by an agency of the US Government. Neither the US Government nor any agency thereof, nor any of their employees, makes any warranty, express or implied, or assumes any legal liability or responsibility for the accuracy, completeness, or usefulness of any information, apparatus, product, or process disclosed, or represents that its use would not infringe privately owned rights. Reference herein to any specific commercial product, process, or service by trade name, trademark, manufacturer, or otherwise does not necessarily constitute or imply its endorsement, recommendation, or favoring by the US Government or any agency thereof. The views and opinions of authors expressed herein do not necessarily state or reflect those of the US Government or any agency thereof.

\end{acknowledgments}

\bibliography{main} 

\end{document}